\documentclass[preprint,nofootinbib]{revtex4}%
\usepackage{amssymb}
\usepackage{amsfonts}
\usepackage{amsmath}
\usepackage{graphicx}%
\begin{document}
\title{Asymptotically anti-de Sitter spacetimes\\in topologically massive gravity}
\author{Marc Henneaux$^{1,2}$, Cristi\'{a}n Mart\'{\i}nez$^{1,3}$, Ricardo
Troncoso$^{1,3}$}
\affiliation{$^{1}$Centro de Estudios Cient\'{\i}ficos (CECS), Casilla 1469, Valdivia, Chile }
\affiliation{$^{2}$Physique th\'{e}orique et math\'{e}matique, Universit\'{e} Libre de
Bruxelles and International Solvay Institutes,ULB Campus Plaine C.P.231,
B-1050 Bruxelles, Belgium}
\affiliation{$^{3}$Centro de Ingenier\'{\i}a de la Innovaci\'{o}n del CECS (CIN), Valdivia, Chile}
\preprint{CECS-PHY-09/02 }

\begin{abstract}
We consider asymptotically anti-de Sitter spacetimes in three-dimensional
topologically massive gravity with a negative cosmological constant, for all
values of the mass parameter $\mu$ ($\mu\neq0$). We provide consistent
boundary conditions that accommodate the recent solutions considered in the
literature, which may have a slower fall-off than the one relevant for General
Relativity. These conditions are such that the asymptotic symmetry is in all
cases the conformal group, in the sense that they are invariant under
asymptotic conformal transformations and that the corresponding Virasoro
generators are finite. It is found in particular that at the chiral point
$|\mu l|=1$ (where $l$ is the anti-de Sitter radius), one must allow for
logarithmic terms (absent for General Relativity) in the asymptotic behavior
of the metric in order to accommodate the new solutions present in
topologically massive gravity, and that these logarithmic terms make
\emph{both} sets of Virasoro generators nonzero even though one of the
central charges vanishes.

\end{abstract}
\maketitle

%\pacs{xxx04.50.+h, yyy04.20.Jb, zzz04.90.+e}

\section{Introduction}

Following the lead of \cite{LSStrominger}, topologically massive gravity with
a negative cosmological constant \cite{Deser:1982vy} has received a great deal
of renewed interest in the last year. The theory is described by the action
\cite{Footnote}
\begin{equation}
I[e]   =2\int\ \left[   e^{a}\left(  d\omega_{a}+\frac{1}{2}\epsilon_{abc}%
\omega^{b}\omega^{c}\right)  +\frac{1}{6}\frac{1}{\ell^{2}}\epsilon_{abc}%
e^{a}e^{b}e^{c} \right]
  +\frac{1}{\mu}\int \omega^{a}\left(  d\omega_{a}+\frac{1}{3}%
\epsilon_{abc}\omega^{b}\omega^{c}\right)  \label{actionTMG}%
\end{equation}
so that the field equations read%
\begin{equation}
G^{\mu}{}_{\sigma}-\frac{1}{l^{2}}\delta_{\sigma}^{\mu}-\frac{1}{\mu}C^{\mu}%
{}_{\sigma}=0,\label{eom}%
\end{equation}
where $\mu\neq0$ is the mass parameter, $l$ is the AdS radius, and $C^{\mu}%
{}_{\sigma}:=\epsilon^{\mu\nu\rho}\nabla_{\nu}\left(  R_{\rho\sigma}-\frac
{1}{4}g_{\rho\sigma}R\right)  $, stands for the Cotton tensor. The case $|\mu
l|=1$ is known as the chiral point and has been advocated in
\cite{LSStrominger} to enjoy remarkable properties.

In the absence of the topological mass term, the relevant asymptotic behavior
of the metric is given by \cite{Brown-Henneaux}
\begin{equation}%
\begin{array}
[c]{lll}%
\Delta g_{rr} & = & f_{rr}r^{-4}+O(r^{-5})\;,\\[2mm]%
\Delta g_{rm} & = & f_{rm}r^{-3}+O(r^{-4})\;,\\[1mm]%
\Delta g_{mn} & = & f_{mn}+O(r^{-1})\;.
\end{array}
\label{Standard-Asympt}%
\end{equation}
Here $f_{\mu\nu}=f_{\mu\nu}(t,\phi)$, and the indices have been split as
$\mu=(r,m)$, where $m$ includes the time and the angle. We have also
decomposed the metric as $g_{\mu\nu}=\bar{g}_{\mu\nu}+\Delta g_{\mu\nu} $,
where $\Delta g_{\mu\nu}$ is the deviation from the AdS metric,
\begin{equation}
d\bar{s}^{2}=-(1+r^{2}/l^{2})dt^{2}+(1+r^{2}/l^{2})^{-1}dr^{2}+r^{2}d\phi
^{2}\;.
\end{equation}
The boundary conditions (\ref{Standard-Asympt}) fulfill the following three
crucial consistency requirements explicitly spelled out in
\cite{Henneaux-Teitelboim}:

\begin{itemize}
\item They are invariant under the anti-de Sitter group.

\item  They decay sufficiently slowly to the exact anti-de Sitter metric at
infinity so as to contain the \textquotedblleft asymptotically anti-de Sitter"
solutions of the theory of physical interest (in this case, the BTZ black
holes \cite{BTZ}).

\item  But at the same time, the fall-off is sufficiently fast so as to yield
finite charges.
\end{itemize}

It was actually found in \cite{Brown-Henneaux} that the asymptotic conditions
(\ref{Standard-Asympt}) are invariant not just under $SO(2,2)$ but under the
bigger infinite-dimensional conformal group in two dimensions. The Poisson
brackets algebra of the corresponding charges (given by surface integrals at
infinity) gives two copies of the Virasoro algebra with a central charge equal
to $c=3l/(2G)$.

If one changes the theory, the asymptotic behavior of the physically
interesting solutions might be different and the asymptotic conditions might
therefore have to be modified in order to accommodate the new solutions of
physical interest. This was investigated at length in
\cite{Henneaux:2002wm,HMTZ2,HMTZ3} for anti-de Sitter gravity with scalar fields in
any number of dimensions (see also \cite{Hertog-Maeda,Marolf}). It was found
that the standard anti-de Sitter boundary conditions indeed had to be relaxed
in that case, but that the charges remained finite thanks to a delicate
cancellation of divergences between the relaxed terms in the metric and
contributions from the scalar fields.

The same phenomenon occurs if one modifies the action of pure Einstein gravity
by the topological mass term, as in (\ref{actionTMG}) above. Indeed, as
observed in \cite{Grumiller-Johansson-Asympt}, the metric could acquire then a
slower decay to the anti-de Sitter metric at infinity for a class of
physically interesting linearized solutions. For a generic value of $\mu l>-1
$, an exact asymptotically AdS solution describing a chiral pp-wave was found
in \cite{DS}, and further developed in \cite{OST}, whose metric reads
\begin{equation}
ds^{2}=l^{2}\frac{dr^{2}}{r^{2}}-r^{2}dx^{+}dx^{-}+F(x^{-})r^{1-\mu l}\left(
dx^{-}\right)  ^{2}\ , \label{pp-wave-mu}%
\end{equation}
where $F(x^{-})$ is an arbitrary function and $x^{\pm}=\frac{t}{l}\pm\phi$.
This solution is to be compared with the AdS metric written in the same
coordinates,
\[
d\bar{s}^{2}=\displaystyle\left(  1+\displaystyle\frac{r^{2}}{l^{2}%
}\right)  ^{-1}dr^{2}-\frac{l^{2}}{4}(dx^{+2}\!+dx^{-2})-\left(
\frac{l^{2}}{2}+r^{2}\right)  dx^{+}dx^{-},
\]
and one sees that the $F(x^{-})r^{1-\mu l}$ term spoils the asymptotic
behavior (\ref{Standard-Asympt}).

The purpose of this note is to provide a consistent set of new boundary
conditions that accommodate these solutions with slower decay at infinity and
that are yet compatible with the full conformal symmetry, for all values of
the mass parameter. It turns out that the analysis carries many features in
common with the scalar case studied previously. (The asymptotic study of
topologically massive gravity has been carried out recently in
\cite{Grumiller-Johansson2} at the chiral point. While we agree with the
asymptotic form of the metric and the symmetries given in that paper, we do
find however that \emph{both} set of Virasoro generators are generically
nonzero, a fact that shows that the theory with these boundary conditions
cannot be chiral \cite{StromingerAugust}.)

Because the computations are rather cumbersome, and because the logic follows
the scalar case situation, we shall, in this note, only report the results and
discuss some of their properties. The full details will be provided elsewhere
\cite{HMTfuture}.

\section{Range $0<|\mu l|<1$ of the mass parameter}

We first consider the most intricate case, which occurs when the mass
parameter $\mu$ fulfills the condition $0<|\mu l|<1$.

\subsection{Asymptotic conditions}

We have found that there are two consistent sets of boundary conditions
fulfilling the three consistency requirements repeated in the introduction.
The existing solutions given in the literature fulfill one or the other set of
boundary conditions. We shall first give the boundary conditions and we shall
then explain how one verifies that they are indeed consistent.

\textit{Negative chirality. }The boundary conditions are in that case
\begin{equation}%
\begin{array}
[c]{lll}%
\Delta g_{rr} & = & f_{rr}r^{-4}+\cdot\cdot\cdot\\
\Delta g_{r+} & = & f_{r+}r^{-3}+\cdot\cdot\cdot\\
\Delta g_{r-} & = & h_{r-}\ r^{-2-\mu l}+f_{r-}r^{-3}+\cdot\cdot\cdot\\
\Delta g_{++} & = & f_{++}+\cdot\cdot\cdot\\
\Delta g_{+-} & = & f_{+-}+\cdot\cdot\cdot\\
\Delta g_{--} & = & h_{--}\ r^{1-\mu l}+f_{--}+\cdot\cdot\cdot
\end{array}
\label{Asympt relaxed metric mu Neg}%
\end{equation}
where $f_{\mu\nu}$ and $h_{\mu\nu}$ depend only on $x^{+}$ and $x^{-}$ and not
on $r$. We use the convention that the $f$-terms are the standard deviations
from AdS already encountered in (\ref{Standard-Asympt}), while the $h$-terms
represent the relaxed terms that need to be included in order to accommodate
the solutions of the topologically massive theory with slower fall-off. We see
that only the negative chirality $h$-terms $h_{r-}$ and $h_{--}$ are present,
hence the terminology.

\textit{Positive chirality. }The boundary conditions are in that case
\begin{equation}%
\begin{array}
[c]{lll}%
\Delta g_{rr} & = & f_{rr}r^{-4}+\cdot\cdot\cdot\\
\Delta g_{r+} & = & h_{r+}\ r^{-2+\mu l}+f_{r+}r^{-3}+\cdot\cdot\cdot\\
\Delta g_{r-} & = & f_{r-}r^{-3}+\cdot\cdot\cdot\\
\Delta g_{++} & = & h_{++}\ r^{1+\mu l}+f_{++}+\cdot\cdot\cdot\\
\Delta g_{+-} & = & f_{+-}+\cdot\cdot\cdot\\
\Delta g_{--} & = & f_{--}+\cdot\cdot\cdot
\end{array}
\label{Asympt relaxed metric mu Pos}%
\end{equation}
with only the positive chirality $h$-terms $h_{r+}$ and $h_{++}$.

Although the known solutions \cite{DS, OST} are of a given chirality and hence
completely covered by the above boundary conditions, one might try to be more
general and include both chiralities simultaneously. This cannot be done,
however, in a manner that is compatible with the other consistency
requirements as it will be explained below.

\subsection{Asymptotic symmetry}

One easily verifies that both sets of asymptotic conditions are invariant
under diffeomorphisms that behave at infinity as
\begin{align}
\eta^{+}  &  =T^{+}+\frac{l^{2}}{2r^{2}}\partial_{-}^{2}T^{-}+\cdot\cdot
\cdot\nonumber\\
\eta^{-}  &  =T^{-}+\frac{l^{2}}{2r^{2}}\partial_{+}^{2}T^{+}+\cdot\cdot
\cdot\label{Asympt KV}\\
\eta^{r}  &  =-\frac{r}{2}\left(  \partial_{+}T^{+}+\partial_{-}T^{-}\right)
+\cdot\cdot\cdot\nonumber
\end{align}
where $T^{\pm}=T^{\pm}(x^{\pm})$. The $\cdots$ terms are of lowest order and
do not contribute to the surface integrals. Hence, the boundary conditions are
invariant under the full conformal group in two dimensions, generated by
$T^{+}(x^{+})$ and $T^{-}(x^{-})$.

\subsection{Surface integrals}

We shall compute the conserved (Virasoro) charges within the canonical
formalism, \textquotedblleft\`{a} la Regge-Teitelboim" \cite{Regge-Teitelboim}%
. The canonical analysis of topologically massive gravity has been performed
in \cite{Deser-Xiang, Carlip}. As noticed in
\cite{Giacomini-Troncoso-Willison}, there is a useful choice of variables
allowing  one to write topologically massive gravity with a cosmological constant as
a Chern-Simons theory. In this case, since the action is already written in
first order, the Hamiltonian formalism can be readily done once the torsion
constraint is incorporated as an additional constraint \cite{Carlip}. This
enables one to skip the standard and somewhat awkward procedure associated
with higher order derivatives.

The charges that generate the diffeomorphisms (\ref{Asympt KV}) take the form
\cite{Regge-Teitelboim}
\begin{equation}
H[\eta]=\hbox{``Bulk piece"}+Q_{+}[T^{+}]+Q_{-}[T^{-}] \,, \label{generator}%
\end{equation}
where the bulk piece is a linear combinations of the constraints with
coefficients involving $\eta^{+},\eta^{-}$, and $\eta^{r}$, which has been explicitly
worked out in \cite{Carlip}, and where $Q_{+}[T^{+}]$ and $Q_{-}[T^{-}]$ are
surface integrals at infinity that involve only the asymptotic form of the
vector field $\eta^{+},\eta^{-}$, and $\eta^{r}$. On shell, the bulk piece vanishes
and $H[\eta]$ reduces to $Q_{+}[T^{+}]+Q_{-}[T^{-}]$.

Evaluating the variation of the bulk piece in (\ref{generator}) under the
asymptotic conditions (\ref{Asympt relaxed metric mu Neg}) or
(\ref{Asympt relaxed metric mu Pos}), one obtains that the surface terms at
infinity should obey:%
\[
\delta Q_{\pm}[T^{\pm}]=\left(  1\pm\frac{1}{\mu l}\right)  \delta Q_{\pm}%
^{0}[T^{\pm}] \,,
\]
where
\[
\delta Q_{\pm}^{0}[T^{\pm}]:=\frac{2}{l} \int T^{\pm}\delta f_{\pm\pm}%
d\phi\ ,
\]
is exactly the same expression as that valid for the standard asymptotic
behavior. The Virasoro charges are then easily integrated to yield
\begin{equation}
\label{once}Q_{\pm}[T^{\pm}]=\frac{2}{l}\left(  1\pm\frac{1}{\mu l}\right)
\int T^{\pm}f_{\pm\pm}d\phi\
\end{equation}
(up to additive constants). The details will be given in \cite{HMTfuture}.
What happens is that the diverging pieces associated with the slower fall-off
$h_{--}$ or $h_{++}$ disappear in $\delta Q_{\pm}[T^{\pm}]$ so that $Q_{\pm}$
is given by (\ref{once}), and hence the charges acquire no correction
involving the terms associated with the relaxed behavior. One can then view
$h_{--}$ (or $h_{++}$), which cannot be gauged away, as defining a kind of
\textquotedblleft hair."  This situation is analogous to the one found for a
scalar field with mass $m$ in the range $m_{ \rm BF}^{2}<m^{2}<m_{\rm BF}^{2}+1/l^{2}$
(where $m_{\rm BF}$ is the Breitenlohner-Freedman bound \cite{BF}). There are then two
possible admissible behaviors (two \textquotedblleft branches") for the
scalar field, and the analysis proceeds as here when only the branch with
slower behavior is switched on \cite{HMTZ3}\footnote{In the scalar field
case, one can switch on simultaneously the two branches in a manner compatible
with AdS asymptotics \cite{Henneaux:2002wm,HMTZ2,HMTZ3,Hertog-Maeda}. If one tries to do this
here, i.e., allows both positive and negative chiralities simultaneously, one
finds that the charges are integrable only if there is a relationship
$h_{++}=h_{++}(h_{--})$ between the two chiralities. This is analogous to what
happens for the scalar field. Contrary to the scalar field case, however, no
such relationship $h_{++}=h_{++}(h_{--})$ is preserved by both the right and
left copies of the Virasoro algebra. Depending on how one chooses the
relationship $h_{++}=h_{++}(h_{--})$, the asymptotic symmetry is reduced to
one copy of the Virasoro algebra (right or left) times $L_{0}$ (left or
right). This is the difficulty mentioned above in trying to include
simultaneously both chiralities. The details will be given in \cite{HMTfuture}%
.}.

Under an asymptotic conformal transformation (\ref{Asympt KV}), $f_{++}$ and
$f_{--}$ are straightforwardly found to transform as
\begin{align}
\delta_{\eta}f_{++}  &  =2f_{++}\partial_{+}T^{+}+T^{-}\partial_{-}%
f_{++}+T^{+}\partial_{+}f_{++}
  -l^{2}\left(  \partial_{+}T^{+}+\partial_{+}^{3}T^{+}\right)
\!/2\ ,\label{deltaf++}\\
\delta_{\eta}f_{--}  &  =2f_{--}\partial_{-}T^{-}+T^{-}\partial_{-}%
f_{--}+T^{+}\partial_{+}f_{--}
  -l^{2}\left(  \partial_{-}T^{-}+\partial_{-}^{3}T^{-}\right)  \!/2\ .
\label{deltaf--}%
\end{align}
On shell, one verifies that
\begin{equation}
\partial_{+}f_{--}=0=\partial_{-}f_{++}%
\end{equation}
and so (\ref{deltaf++}) and (\ref{deltaf--}) reduce to
\begin{align}
\delta_{\eta}f_{++}  &  =2f_{++}\partial_{+}T^{+}+T^{+}\partial_{+}%
f_{++}-\frac{l^{2}}{2}\left(  \partial_{+}T^{+}+\partial_{+}^{3}T^{+}\right)
,\label{delta2h++}\\
\delta_{\eta}f_{--}  &  =2f_{--}\partial_{-}T^{-}+T^{-}\partial_{-}%
f_{--}-\frac{l^{2}}{2}\left(  \partial_{-}T^{-}+\partial_{-}^{3}T^{-}\right)
. \label{delta2h--}%
\end{align}
As $\beta_{+}\,\delta_{\eta}\int f_{++}Y^{+}d\phi\sim\lbrack Q_{+}%
(Y^{+}),Q_{+}(T^{+})+Q_{-}(T^{-})]$ (with $\beta_{\pm}=2l^{-1}\left(  1\pm(\mu
l)^{-1}\right)  $) and $\beta_{-}\,\delta_{\eta}\int f_{--}Y^{-}d\phi
\sim\lbrack Q_{+}(Y^{+}),Q_{+}(T^{+})+Q_{-}(T^{-})]$ (with $Y^{+}$ and $Y^{-}$
the asymptotic conformal transformation associated with a second spacetime
diffeomorphism $\xi^{+}$, $\xi^{-}$ and $\xi^{r}$), one can easily infer from
(\ref{delta2h++}) and (\ref{delta2h--}) that $Q_{+}(Y^{+})$ and $Q_{-}(T^{-})$
commute with each other and each fulfills the Virasoro algebra with central
charges
\begin{equation}
c_{\pm}=\left(  1\pm\frac{1}{\mu l}\right)  \,c
\end{equation}
(see \cite{Brown-Henneaux2} for general theorems).

\section{Range $|\mu l|>1$ of the mass parameter}

Take for definiteness $\mu l$ positive and hence $>1$. Solving the equations
starting from infinity shows that again, one should expect both chiralities to
be present, taking exactly the same form as
(\ref{Asympt relaxed metric mu Neg}) and (\ref{Asympt relaxed metric mu Pos})
above. However, the positive chirality blows up at infinity ($\Delta g_{++}$
dominates the background) and the space is not asymptotically of constant
curvature. So, if $h_{++}\not =0$, the space is not asymptotically anti-de
Sitter. For this reason, one must set $h_{++}=0$. But the other $h_{--}$-term
is subdominant with respect to $f_{--}$, so that the asymptotic negative
chirality behavior reproduces (\ref{Standard-Asympt}). The same analysis
holds when $\mu l$ is negative (with an interchange of the roles of the two
chiralities). Therefore, the behavior of the metric can be taken to be
(\ref{Standard-Asympt}). The asymptotic derivation of the charges and the
central charges proceeds then straightforwardly (no divergence to be canceled)
and yields
\begin{equation}
Q_{\pm}[T^{\pm}]=\frac{2}{l}\left(  1\pm\frac{1}{\mu l}\right)  \int T^{\pm
}f_{\pm\pm}d\phi\
\end{equation}
with central charges
\begin{equation}
c_{\pm}=\left(  1\pm\frac{1}{\mu l}\right)  \,c\ .
\end{equation}

\section{The chiral point}

\subsection{Asymptotic behavior}

Hereafter we only consider $\mu l=1$, since the case of $\mu l=-1$ just
corresponds to the interchange $x^{+}\longleftrightarrow x^{-}$.

In the case of $\mu l=1$, the appropriate asymptotic behavior for $\Delta
g_{\mu\nu}$ reads%

\begin{equation}%
\begin{array}
[c]{lll}%
\Delta g_{rr} & = & f_{rr}r^{-4}+\cdot\cdot\cdot\\
\Delta g_{r+} & = & f_{r+}r^{-3}+\cdot\cdot\cdot\\
\Delta g_{r-} & = & h_{r-}\ r^{-3}\ln\left(  r\right)  +f_{r-}r^{-3}%
+\cdot\cdot\cdot\\
\Delta g_{++} & = & f_{++}+\cdot\cdot\cdot\\
\Delta g_{+-} & = & f_{+-}+\cdot\cdot\cdot\\
\Delta g_{--} & = & h_{--}\;\ln\left(  r\right)  +f_{--}+\cdot\cdot\cdot
\end{array}
\label{Asympt relaxed metric}%
\end{equation}
where $f_{\mu\nu}$ and $h_{--}$ depend only on $x^{\pm}=\frac{t}{l}\pm\phi$.
This behavior accommodates the known solutions with constant curvature at
infinity \cite{DS, OST,Gaston}, whose metric is given by
\begin{equation}
ds^{2}=l^{2}\frac{dr^{2}}{r^{2}}-r^{2}dx^{+}dx^{-}+F(x^{-})\log(r)\left(
dx^{-}\right)  ^{2}\ . \label{pp-wave chiral}%
\end{equation}
with $F(x^{-})$ being an arbitrary function.

\subsection{Asymptotic symmetry}

Just as for $\mu l\not =1$, the asymptotic conditions are invariant under
diffeomorphisms that behave at infinity as in Eq. (\ref{Asympt KV}), where the
$\cdots$ terms are again of lowest order and do not contribute to the surface
integrals. Hence, the boundary conditions are invariant under the conformal
group in two dimensions, generated by $T^{+}(x^{+})$ and $T^{-}(x^{-})$.

Under the action of the Virasoro symmetry, one obtains%
\begin{equation}
\delta_{\eta}h_{--}=2h_{--}\partial_{-}T^{-}+T^{-}\partial_{-}h_{--}%
+T^{+}\partial_{+}h_{--}\ \label{deltah--chiral}%
\end{equation}
and
\begin{equation}
\delta_{\eta}f_{++}   =2f_{++}\partial_{+}T^{+}+T^{-}\partial_{-}%
f_{++}+T^{+}\partial_{+}f_{++}
 -l^{2}\left(  \partial_{+}T^{+}+\partial_{+}^{3}T^{+}\right)  \! /2 .
\label{deltaf--chiral}%
\end{equation}

The field equations are easily verified to imply that%
\[
\partial_{-}f_{++}=0\text{, and }\partial_{+}h_{--}=0\text{.}%
\]
Note that this time the equations do not impose $\partial_{+}f_{--}=0$ and
furthermore, the transformation rule of $f_{--}$ also differs from the one
found off the chiral point.

\subsection{Conserved \textbf{charges for }$\mu l=1$}

Evaluating the variation of the surface charges using the expressions of
\cite{Carlip} with the asymptotic conditions (\ref{Asympt relaxed metric}) one
obtains:%
\[
\delta Q_{+}=\frac{4}{l}\int T^{+}\delta f_{++}d\phi\ ,\text{ and }\delta
Q_{-}=\frac{2}{l}\int T^{-}\delta h_{--}d\phi\ .
\]
This implies (up to additive constants)
\[
Q_{+}=\frac{4}{l}\int T^{+}f_{++}d\phi\ ,\text{ and }Q_{-}=\frac{2}{l}\int
T^{-}h_{--}d\phi\ .
\]
The crucial new feature found here, apparently overlooked in the previous
literature, is that $Q_{-}[T^{-}]$ does not vanish identically. Rather, the
relaxation term $h_{--}$ does contribute to it. This behavior is somehow
similar to what occurs for scalar fields that saturates the BF bound. One may
verify explicitly that on definite solutions, $Q_{-}[T^{-}]$ is not zero
\cite{HMTfuture}. Indeed, for the metric (\ref{pp-wave chiral}),
$h_{--}=F(x^{-})$ which in general does not vanish.

{}From the variations (\ref{deltah--chiral}) and (\ref{deltaf--chiral}) of
$h_{--}$ and $f_{++}$ and the asymptotic field equations, one finds that both
$Q_{+}[T^{+}]$ and $Q_{-}[T^{-}]$ fulfill the Virasoro algebra with the
central charge
\begin{equation}
c_{+}=2\,c\,,\;\;\;\;c_{-}=0\,.
\end{equation}
Even though $Q_{-}[T^{-}]$ does not vanish, the central charge $c_{-}$ is zero
because the inhomogeneous terms $-l^{2}\left(  \partial_{-}T^{-}+\partial
_{-}^{3}T^{-}\right)  /2$ are absent from $\delta_{\eta}h_{--}$.

In this paper we have exhibited the boundary conditions appropriate to
accommodate the solutions of topologically massive gravity found in the
literature with a slower decay at infinity than the one for pure standard
gravity discussed in \cite{Brown-Henneaux}. These boundary conditions fulfill
the consistency conditions listed in the introduction. The analysis proceeds
very much as in the case of anti-de Sitter gravity coupled to a scalar field
\cite{Henneaux:2002wm,HMTZ2,HMTZ3} and the results turn out to be comparable.

A question not addressed here is the exact physical relevance of the new
solutions which the more liberal boundary conditions enable one to consider.
One might question whether they should be included \cite{StromingerAugust} and
it is not clear what one loses if one does not include them, i.e., if one
sticks to the more restrictive boundary conditions of \cite{Brown-Henneaux}.
Note that for the scalar field, the softening of the boundary conditions leads
to interesting developments. In particular, this enlarges the space of
admissible solutions to include hairy black holes
\cite{Henneaux:2002wm,Hertog-Maeda,MTZ}, solitons and instantons \cite{GMT}.

We have also shown that with the new boundary conditions, the Virasoro
generators with both chiralities are actually nonzero at the chiral point
(while one chiral set of them does vanish under the boundary conditions of
\cite{Brown-Henneaux}). The corresponding central charge vanishes, however.
This puzzling fact should be understood from the point of view of conformal
field theory.

A longer version of this paper, with detailed proofs and more information on
the charges of various solutions is in preparation \cite{HMTfuture}.

\emph{Note added}: After this paper was posted on the arXiv, we received
comments by various colleagues (i) confirming the intriguing result
established above that at the chiral point, the left-moving generators
have a zero central charge even though they are not identically zero;
and (ii) investigating the interpretation of this result in terms of a
dual logarithmic CFT.  We thank A. Strominger and D. Grumiller for
kindly providing us this information prior to publication

\textit{Acknowledgments.} We thank G. Comp\`{e}re, G. Giribet, D. Grumiller,
N. Johansson, A. Schwimmer and A. Strominger for useful discussions and enlightening comments.
This research is partially funded by FONDECYT grants 1061291,
1071125, 1085322, 7080044, 1095098. The work of MH is partially
supported by IISN - Belgium (conventions 4.4511.06 and 4.4514.08) and by the
Belgian Federal Science Policy Office through the Interuniversity Attraction
Pole P6/11. C. M. and R. T. wish to thank the kind hospitality at the Physique
th\'{e}orique et math\'{e}matique at the Universit\'{e} Libre de Bruxelles and
the International Solvay Institutes. The Centro de Estudios Cient\'{\i}ficos
(CECS) is funded by the Chilean Government through the Millennium Science
Initiative and the Centers of Excellence Base Financing Program of Conicyt.
CECS is also supported by a group of private companies which at present
includes Antofagasta Minerals, Arauco, Empresas CMPC, Indura, Naviera Ultragas
and Telef\'{o}nica del Sur. CIN is funded by Conicyt and the Gobierno Regional
de Los R\'{\i}os.


\begin{thebibliography}{99}
\bibitem {Footnote}Here we are using the conventions of \cite{Carlip}: $16\pi
G=1$, a metric of signature $(-++)$, and a cosmological constant
$\Lambda=-1/l^{2}$, and set $\epsilon_{012}=1$ (so $\epsilon^{012}=-1$).
\end{thebibliography}
\end{document}